\documentstyle[rotating,12pt,titlepage,epsfig,a4,amssymb]{article}
\newcommand{\pom}{{\mathbb{P}}}

\newcommand{\xpom}{x_\pom}

\begin{document}
\begin{titlepage}
  
\begin{center}

\vspace*{1cm}

\Large
\bf
            The H1 Forward Proton Spectrometer at HERA

\rm            
\large
\vspace*{1cm}
P.van Esch \\
{\small Inter-University Institute for High Energies ULB-VUB,
  Brussels, Belgium}
\vspace*{5mm}

M.Kapichine, A.Morozov, V.Spaskov \\
{\small Joint Institute of Nuclear Research, Dubna, Russia} 
\vspace*{5mm}

W.Bartel, B.List\footnotemark[1], H.Mahlke-Kr\"uger, V.Schr\"oder,
T.Wilksen \\
{\small DESY, Hamburg, Germany} 
\footnotetext[1]{Presently research associate at CERN.}
\vspace*{5mm}

F.W.B\"usser, K.Geske, O.Karschnik, F.Niebergall, H.Riege, J.Sch\"utt,
R.van Staa, C.Wittek \\
{\small II. Institut f\"ur Experimentalphysik, Universit\"at Hamburg,
  Hamburg, Germany} 
\vspace*{5mm}

D.Dau \\
{\small Institut f\"ur Reine und Angewandte Kernphysik, Universit\"at
  Kiel, Kiel, Germany}
\vspace*{5mm}

D.Newton \\
{\small School of Physics and Chemistry, University of Lancaster,
  Lancaster, UK}
\vspace*{5mm}

S.K.Kotelnikov, A.Lebedev, S.Rusakov \\
{\small Lebedev Physical Institute, Moscow, Russia} 
\vspace*{5mm}

A.Astvatsatourov, J.B\"ahr, U.Harder, K.Hiller\footnotemark[2],
B.Hoffmann\footnotemark[3], H.L\"udecke, R.Nahnhauer \\
{\small DESY Zeuthen, Zeuthen, Germany}
\vspace*{5mm}

\footnotetext[2]{Corresponding author, Tel.: +49 33762 77208, Fax:
 +49 33762 77330, e-mail: hiller@ifh.de }
\footnotetext[3]{Now Berlin Chemie AG, 12489 Berlin, Glienicker Weg 125.}

\end{center}
\end{titlepage}

\vspace*{5cm}
\large
\begin{abstract}
\vspace*{2cm}
\normalsize

The forward proton spectrometer is part of the H1 detector at the HERA
collider. Protons with energies above 500 GeV and polar angles below
1 mrad can be detected by this spectrometer. The main detector
components are scintillating fiber detectors read out by
position-sensitive photo-multipliers. These detectors are housed in
so-called Roman Pots which allow them to be moved close to the
circulating proton beam. Four Roman Pot stations are located at
distances between 60 m and 90 m from the interaction point.

\vspace*{1cm}
\noindent
\it {PACS:} \hspace*{6mm} \rm 29.30.Aj, 29.40.Gx 

\end{abstract}

\normalsize

\section{Introduction}
The forward proton spectrometer (FPS) is part of the H1 detector at
the HERA collider at DESY [1]. From 1995, the first year of FPS
operation, until 1997 HERA collided 27.5~GeV positrons
with 820 GeV protons. In 1998 positrons were replaced by
electrons~\footnote{Throughout this paper electron is a generic name
for $e^{-}$ and $e^{+}$.}
and the proton energy was increased to 920 GeV.
 
The aim of the FPS is to extend the acceptance of the H1 detector in
the very forward direction, which is the direction of the outgoing
proton beam. The central tracking chambers cover the angular range
down to polar angles of $5^{o}$. At smaller angles energy deposits are
observed in detectors close to the outgoing proton beam direction. The
liquid argon calorimeter, the plug calorimeter and the proton remnant
tagger in combination with the forward muon system are sensitive to
forward activities down to polar angles of about 1 mrad [2]. The FPS
has been built to measure forward going protons which are scattered at
polar angles below this range.

Forward going protons with energies close to the kinematic limit of
the incident proton beam energy arise from
diffractive processes, which in Regge theory are described by Pomeron
exchange [3]. At lower proton energies processes with meson exchange
become the dominant production mechanism [4]. The measurement of
diffractive processes offers the possibility to investigate the
structure of the exchanges [5,6].

The FPS measures the trajectory of protons emerging from
electron-proton collisions at the interaction point. The HERA magnets
in the forward beam line separate scattered protons from the
circulating proton beam. After a distance of about 60 m the scattered
protons deviate typically a few millimeters from the central proton
orbit and can be measured by tracking detectors.

These detectors are multi-layer scintillating fiber detectors read out
by position-sensi-\\
tive photo-multipliers (PSPMs). Scintillator tiles
cover the sensitive detector area and are used to form trigger
signals. To provide the necessary aperture for the proton injection
into HERA
the detectors are mounted in movable plunger vessels, so-called Roman
Pots, which are retracted during the beam filling and orbit tuning
process [7]. When stable conditions are reached the detectors are
moved close to the cirulating beam.

Due to the optics and the construction of the beam line only one
secondary particle with at least 500 GeV energy can reach the FPS
stations. The track parameters of the measured trajectories are used to
reconstruct the energy and the scattering angles at the interaction
point based on the knowledge of bending strengths and positions of the
HERA magnets.

Four FPS stations at distances between 64 m and 90 m from the
interaction point measure the trajectory of scattered protons. In 1995
two vertical stations were built, which approach the beam from above.
They detect protons in the kinematic range $0.5 <~E'/E_p < 0.95$,
where $E'$ and $E_p$ are the energies of the scattered and the
incident protons, respectively. In 1997 two horizontal stations were
added, which approach the beam from the outer side of the proton ring.
Complementary to the vertical FPS stations, they are sensitive in the
range $E'/E_p > 0.9$.

This paper is organized as follows: section 2 contains the details of the 
FPS hardware components. In section 3 the operation of the FPS and the
experience after three years of data taking are discussed. The main
detector characteristics as acceptance, efficiency and resolution
are given in section 4. In this section also the method of energy
reconstruction and calibration are described. An summary of the
relevant FPS parameters and an outlook on physics results is given
in section 5.

\section{Detector Components}
The positions of the four FPS stations along the forward beam line are
shown in fig.1. The two vertical stations are placed at 81 m and 90 m
behind the BU00 dipole magnet, which bends the proton beam 5.7 mrad
upwards. In front of this magnet at 64 m the first horizontal station
is placed while the second horizontal station is located behind the
BU00 magnet at 80 m.

\subsection{Mechanics}
All FPS stations consist of two major mechanical components:
the plunger vessel which is the movable housing of the detector
elements, and the fiber detectors with their readout
components.

The plunger vessel is a cylinder made of 3 mm stainless steel. The
inner volume, where the fiber detectors are mounted, has a diameter of
140 mm. To reduce the amount of material in front of the detectors
thin windows of 0.3 mm steel are welded to the main cylinder. The
bottom sections at the positions of the detectors are closed by 0.5 mm
steel plates.
   
The plunger vessel can be moved close to the proton beam orbit due to
the flexible connection via steel bellows to a flange of the beam
pipe. In the vertical stations the plunger vessel is driven by
spindles which are connected by a belt to a stepping motor.
Due to space limitations the horizontal stations have a hydraulic
moving system. The maximum range for the detector movement is 50 mm in
the vertical stations and 35 mm in the horizontal stations.
All stations are equipped with external position measuring devices
\footnote{Messtaster Metro MT60, Dr.J.Heidenhain GmbH, D-8225
  Traunreut.} 
which give the actual detector positions with 10 $\mu$m precision.
These values are recorded every second by the slow control system.

A sketch of the detector insert of the horizontal stations is shown in
fig.2a. The detector carrier is an aluminum tube with a 
carbon fiber end part to support the fiber detectors. The platform
above the detector carrier houses the PSPMs and the front-end
electronics. The trigger photo-multipliers (PMTs) are fixed on the
detector carrier inside the plunger vessel. In the vertical stations
the basic arrangement is the same, but the trigger PMTs are placed
outside the plunger vessel on the same platform as the PSPMs.

For the reconstruction of the proton energy the positions of all FPS
stations with respect to the HERA proton beam have to be known. For
this purpose a measuring plate with four marks is inserted into
the plunger vessel. These marks are geodetically surveyed with a
precision of 100 $\mu$m. The position of the fiber detectors with
respect to the outer marks is measured on a scanner with a precision
of 20 $\mu$m.

The front-end electronics together with the PSPMs and trigger PMTs
has a power consumption of about 150 W. To keep the temperature below
$30^{o}$C the walls of the platform housing the electronic components
are water cooled.

The front-end electronics of all FPS stations is surrounded by an external
lead shielding against synchrotron radiation. The horizontal stations
have in addition a soft iron shielding to reduce the influence of
magnetic stray fields on the PSPMs.

For safety reasons the plunger vessels are filled with dry nitrogen
gas, so that in case of leakage only a small amount of dry gas enters
the HERA machine vacuum.

\subsection{Fiber Detectors}
Scintillating fibers arranged in multi-layer structures represent a
fast and robust tracking detector. The vertex smearing and beam
divergence at the interaction point set the scale for the spatial
detector resolution. Due to these constraints, the energy resolution
cannot be improved by a detector resolution better than 100$\mu$m.

A view of the arrangement of the fiber detectors in the vertical
stations is given in fig.2b. Each pot is equipped with two identical
subdetectors measuring two coordinates transverse to the beam
direction. The subdetectors are separated by 60 mm to allow the
reconstruction of a local track segment of the proton trajectory. Each
subdetector consists of two coordinate detectors with fibers inclined
by $\stackrel{+}{-} 45^{o}$ with respect to the symmetry axis of the
$\pm 45^{o}$ with respect to the symmetry axis of the plunger vessel.
This angle was chosen to avoid a strong bending of the light guide
fibers inside the plunger vessel.

Each coordinate detector consists of five fiber layers to ensure good
spatial resolution and efficiency. The fibers of 1 mm diameter are
positioned in parallel to each other with a pitch of 1.05 mm within each
layer. Neighbouring fiber layers are staggered by 0.21 mm to obtain the 
best spatial resolution.

The size of the fiber detector was chosen to detect most of the
scattered protons in one hemisphere of the proton orbit. According to
Monte Carlo simulations a detector size of about $5 \times 5$ cm$^2$
meets this requirement in the vertical stations, while in the
horizontal stations detectors of half this size are sufficient. Hence
in the vertical stations 48 fibers are combined into one layer while 24
fibers per layer are used in the horizontal stations.

The precision of the fiber positions at the detector end
face was measured by a microscope. The typical deviation from the
nominal fiber position is 10 $\mu$m.

The scintillating fibers are thermally spliced to light guide fibers
which transmit the scintillation signals to the PSPMs. The light
guides have a length of 50 cm in the vertical stations and 30 cm in
the horizontal stations. The light transmission of all spliced
connections was measured and only those with a transmission better
than 80 \% were accepted for the detector production.

To optimize the mechanical stability of the spliced connections the
fibers of several producers were tested. The chosen scintillating
fibers \footnote
{POLIFI 02-42-100, Pol.Hi.Tech.,S.P.Turanense Km.44400 - 67061
Carsoli(AQ),Italy.}
have an attenuation length of 
3.5~m and a trapping efficiency of 4 \%.
In a test run the light yield of 4.5 photo-electrons per millimeter fiber
traversed by a minimum ionizing particle was measured at the end of 
2~m light guides [8]. For the detectors in the horizontal stations
fibers with double cladding were used. Due to a trapping efficiency of
nearly 7 \% these fibers have an enhanced light yield.

The light guide fibers are glued into a plastic mask to feed
the scintillation signals into the PSPM channels. All fibers of a
coordinate detector are combined into the same mask to be read out by
one PSPM.

\subsection{Position-Sensitive Photo-Multipliers}
The position-sensitive photo-multiplier (PSPM) is an efficient device
to detect scintillation signals
from many fibers simultaneously. Different types of PSPMs are
used in the vertical and horizontal stations. For the vertical stations
the 64-channel PSPM H4139-20
\footnote{HAMAMATSU PHOTONICS K.K. Electron Tube Center,314-5
Shimokanzo, 438-0193 Japan.}
was chosen. This device gave the best results for the readout of the
fiber detectors in terms of efficiency and resolution at that time.
For the FPS upgrade the 124-channel PSPM MCPM-124
\footnote{MELZ, Electrozavodskaja 14, 107061 Moscow, Russia.}
was applied in the horizontal stations.

The fundamental difference between the two types of PSPMs is the
electron multiplication system. The H4139-20 has a fine-mesh dynode
system which gives a gain above $10^6$ while the MCPM-124 is equipped
with two micro-channel plates which produce a gain typically one order
of magnitude less. An important feature of the MCPM-124 is the
electro-static focusing and the anti-distortion electrode between the
photo-cathode and the first micro-channel plate. Due to the long path
of the photo-electrons this device is very sensitive to magnetic
fields. A detailed investigation of the micro-channel plate PSPM can
be found in [9]. The main characteristics of both PSPM types are
compiled in table 1.

For the correct recognition of fiber hits the cross talk to
neighbouring pixels should be small. The values quoted by the producer
range around a few percent. However a large contribution of electronic
cross talk can increase the overall cross talk above the level of 
10 \%.

To read out all 240 fibers of a coordinate detector in the vertical
stations by 64 channels of the H4139-20 type implies that 4 fibers
have to be coupled to one PSPM pixel. The pixel size of 4 mm diameter
is large enough to place 4 fibers of 1 mm diameter without increasing
the cross talk. The consequence of this 4-fold fiber multiplexing is a
4-fold ambiguity in the hit recognition.
Since the FPS stations aim to measure only one particle track 
this ambiguity can be resolved by a corresponding
segmentation of the trigger planes into four scintillator tiles
(section 2.4).

In the horizontal stations all 120 fibers of a coordinate detector can
be read out by the 124 channels of the MCPM-124 without multiplexing.
The pixel size of $1.5 \times 1.5$ mm$^2$ matches well the fiber
diameter of 1 mm. The distortions of the regular anode pixel grid due
to the electro-static focusing are compensated by an appropriate
design of the fiber mask. 

A general principle of the fiber-to-PSPM-pixel mapping is, that
neighbouring fibers are not read out by neighbouring PSPM pixels.
This scheme reduces the influence of cross talk on the spatial
resolution of the detector.

All PSPMs have 2 pixels which are coupled via light guide fibers to
light emitting diodes for monitoring and test purposes.

\subsection{Trigger Counters}
The trigger counters consist of scintillator planes covering the
sensitive area of the fiber detectors. Each coordinate detector is
equipped with one scintillator plane, placed before or behind the
fiber detector. Altogether there are four scintillator planes in each
FPS station.

The arrangement of the scintillator planes with respect
to the fiber detectors in the vertical stations can be seen in fig.2b.
To resolve the spatial ambiguity due to the 4-fold fiber multiplexing
the scintillator planes are segmented into four tiles. Each tile
covers 12 fibers with a unique fiber-to-PSPM pixel mapping.
The trigger tiles are made of 5 mm thick scintillator material BC-408
\footnote{BICRON, 12345 Kinsman Road, Newbury, Ohio 44065-9577,USA.}.
Bundles of 240 light guide fibers of 0.5 mm diameter are used to
transmit the light signals to the trigger PMTs. The bundles are glued
to the end faces of the scintillator tiles and have a length of 50 cm.
Altogether 16 PMTs XP1911~\footnote{Philips Components, Postbus 90050, 5600 PB Eindhoven, Netherlands.}
are used to read out the trigger signals in a vertical station.

In the horizontal stations the scintillator planes are not segmented
and only 3 mm thick. Each plane is connected to two bundles of light
guide fibers to transmit the scintillation signals to two PMTs R5600
\footnote{HAMAMATSU PHOTONICS K.K. Electron Tube Center, 314-5 Shimokanzo,
438-0193 Japan.} - a metal package PMT characterized by compact size and low
weight.
Since in the horizontal stations the trigger PMTs are placed inside
the plunger vessel the light guide bundles have a length of only 15 cm.
The main parameters of the trigger PMTs are summarized in table 2.

\subsection{Electronics}
The main parts of the FPS electronics are given in a block diagram in
fig.3. It can be subdivided into three parts:

\vspace*{5mm}

\noindent
$\bullet$ the front-end components: preamplifiers and comparators for the
   signals of the PSPMs and trigger PMTs mounted close to the detectors; \\
$\bullet$  the conversion, pipelining and trigger electronics located in
   crates a few meters away from the FPS stations in the HERA
   tunnel;\\
$\bullet$  the VME master controller residing outside the HERA tunnel
    organizing the data readout and the trigger processing.\\

In the vertical stations, which are equipped with the H4139-20, one can
expect an anode charge of 0.5 pC per fiber hit. A charge sensitive
preamplifier with a sensitivity of 1 V/pC is used as the first step in
the signal processing chain, followed by a differential driver
circuit for common mode rejection. In the horizontal stations the
preamplifier sensitivity is enlarged to 30 V/pC due to the lower gain
of the MCPM-124. The analog signals of the trigger PMTs are fed into
amplifiers to enlarge their amplitudes by a factor of 10 before the
digitization.

The first stage of the signal processing is a FADC with 6 bit
resolution and 1 V range. It is strobed by the HERA clock and the
digitized signals are packed into 8 bit wide pipeline registers.
Since HERA has a bunch crossing interval of 96 nsec and the first
level trigger decision from the main H1 detector is available only
after 2.5 $\mu$sec all data have to be stored in pipelines. The FPS
pipeline boards have a length of 32 bunch crossings.

In addition to the FADC readout the signals of the trigger PMTs
are fed into comparator boards with a remotely controlled threshold.
The  output signals are transmitted to the trigger board which compares
the hit pattern with trigger conditions consistent with a proton
track segment. The logical OR of all conditions comprises the local
trigger signal of a FPS station. It is stored together with the hit
pattern of all trigger PMTs on the pipeline board.

All local trigger signals are combined to FPS trigger elements
for the central H1 trigger processor. To compensate the different
distances of the FPS stations to the interaction point programmable
delay circuits are used to synchronize the FPS trigger pulses with the
HERA clock.

The FPS data are transmitted via a bi-directional fiber optic link to
the input FIFO of the VME master controllers. All FPS stations are read
out in parallel with a strobe frequency of 5 MHz in a maximum readout
time of 52 $\mu$sec.

A second fiber optic link is used to transmit the signals from the
trigger boards to the VME master controllers outside the HERA tunnel.
In the other direction the HERA clock signals and information from the
central H1 trigger unit are transmitted to each FPS station.

\section{Operation and Data Taking}
\subsection{Operation of the FPS}
The beam profile determines how close the detectors can be moved
to the proton beam orbit. It can be calculated from the
emittance and the $\beta$-function of the proton machine [10].
At the positions of the FPS stations the width and the height of the
beam profile are given in table 3. It is a flat ellipse at 90 m and
80.5 m and becomes wider at 64 m.

Assuming a distance of 10 standard deviations of the beam profile
plus a safety margin of 2 mm as the closest distance of approach, the
detectors in the vertical stations can be moved up to 4 mm to the
central proton orbit. In the horizontal stations the distance of
closest approach is about 30 mm at 64 m and 20 mm at 80m.

During injection and ramping of the beams all FPS detectors are in
their parking positions far from the circulating beam. When stable
beam conditions are reached an automatic insert procedure is started.
The detectors are moved in steps of 100 $\mu$m and the increase
of counting rates in the trigger tiles and the beam loss monitors
mounted on the beam pipe at 64 m, 83 m and 95 m are observed. For the
vertical station at 90 m also the rate of the forward neutron 
calorimeter [11] located at a distance of 107 m from the
interaction point is recorded.

As the detectors are moved towards the proton beam a gradual increase
of the rates is observed as long
as the plunger vessels remain in the shadow of the HERA machine
collimators which cut the tails of the proton beam profile. When the
plunger vessel leaves the collimator shadow the rates increase
steeply.

The automatic insert procedure evaluates the gradient of the 
rate increase to stop the movement. If the increase between two
consecutive steps exceeds a predefined level the movement is stopped
and the detectors have reached their working positions.

The insert procedure starts with the station nearest to the
interaction point at 64 m, because particles scattered off the bottom
of the plunger vessel affect the rates in the FPS stations further
downstream. The whole procedure until all detectors have reached their
working positions takes about 20 minutes.

The slow control system records detector parameters in
time intervals of seconds and stores them into the database. Some
important parameters are the detector positions, the values of the
HERA beam position monitors and the trigger rates.

The dose rate in all FPS stations accumulated over one year of HERA
operation was measured. At the positions of the fiber detectors the
highest rate was 28 krad in the horizontal station at 64 m. The other
stations are well protected by the yoke of the BU00 dipole magnet and
accumulated doses below 1 krad. The dose rates in all stations are
well below the limits of radiation damage of scintillating fibers [12].

An emergency retract system protects the detectors and PSPMs against
radiation damage by badly tuned beams or accidental beam loss. When
the rate monitors indicate a rate above a critical threshold all
detectors are quickly moved to their parking positions.

An important item for the operation of the FPS is the safety of the
HERA running. The motor drives of all FPS stations are connected to
the H1 emergency power net and in case of a power break all
detectors are automatically retracted. In addition, the horizontal
stations have a spring loaded sytem to retract the detectors in case
of problems. During the detector insert procedure several checks
protect the proton beam against accidental detector movement by human
intervention or computer failures.

\subsection{Data Acquisition Program}
The data acquisition (DAQ) has two major tasks: it organizes
the data transfer from the
front-end electronics to the central H1 event builder and monitors
data quality parameters. 

When a first level trigger decision stops the filling of the pipeline
a controller program starts the readout of the FADC data of a specific
set of pipeline stages, which includes all PSPM and trigger PMT
signals. The digital trigger data of five consecutive stages are read
out including the central pipeline stage containing the FADC data.
Due to the long distance to the central H1 detector the signals of a
proton traversing the FPS stations reside close to the end of the
pipeline. 

The master controllers receive the data via a fiber-optical link
and perform a pedestal subtraction with zero suppression. Finally, the
DAQ program reformats the data for the off-line analysis and sends
them to the central event builder.

The size of a typical FPS event is 200 Byte. About 1 \% to 2 \%
of the total H1 data sample contains such an FPS event. The deadtime
due to the FPS DAQ is typically 
1.2~msec.

The monitor program allows to modify readout parameters
as the comparator thresholds of the trigger PMTs, the definition of
the trigger conditions, the FADC strobe delays and the zero suppression
thresholds of the PSPM amplitudes. In addition it shows control
histograms and an online event display.

\section{Detector Performance}

\subsection{Trigger Rate and Efficiency}
The trigger rate of the scintillator tiles has contributions from two
major sources:
the signals from traversing particles, either scattered protons or
shower particles from beam-gas or beam-wall interactions, and the
background rate due to PMT noise and synchrotron radiation from the
electron machine. To reduce the synchrotron background rate 
all stations are shielded with lead plates.

The majority of the data were recorded with the trigger condition that
at least 3 out of 4 trigger planes per station have fired. Moreover,
in the vertical stations the combination of scintillator tiles which
correspond to a track topology define narrow forward cones. In the
horizontal stations the unsegmented scintillator planes are coupled to
two PMTs which are used in coincidence to reduce the contribution
of PMT noise.

The resulting trigger rates of the vertical FPS stations are between 2
kHz and 4~kHz for typical luminosity conditions with proton and electron
currents of 60 mA and 20~mA, respectively. The coincidence rate of the
vertical stations varies between 0.5 kHz and 1.5~kHz. In the
horizontal stations the trigger rates are significantly higher. This
is mostly due to the wide beam profile in the horizontal coordinate
(table 3) and the missing shielding of the huge BU00 dipole magnet
before the station at 64 m. The typical coincidence rate of the
horizontal stations is about 10 kHz.

The four local trigger signals of all FPS stations are combined to
8 FPS trigger elements. They are used in combination with trigger
elements from other H1 subdetectors in the central trigger processor.
Three physics triggers with the signature of a forward going proton
are formed for low multipicity, photo-production and deep inelastic
scattering events.

The efficiency of the trigger tiles is determined from redundant 
signals which can be attributed to the same forward track.
The typical values of the tile efficiency are well above 98 \%.

\subsection{Hit Identification and Track Reconstruction }
The signals of all PSPM channels are classified into hit, noise,
or cross talk signals before the track reconstruction.

To reduce the influence of cross talk on the track reconstruction a
filter algorithm is applied [13]. Possible cross talk signals 
in the neighbourhood of channels with large
amplitudes are suppressed, while isolated channels with small
amplitudes are kept. A signal  above an amplitude
threshold of 2 $\sigma_i$ + 1 FADC counts, where $\sigma_i$ is the
pedestal variation of the i-th PSPM channel, is accepted as a hit
if at least one of the two associated trigger tiles has fired.

In the first step of the track reconstruction the fiber hits in a
coordinate detector are grouped into clusters compatible with a forward
track segment. Hits in at least two layers are requested for each
cluster.

As described in section 2.2 each FPS station contains two identical
subdetectors separated by 60 mm. Each cluster in the first
detector is combined with each cluster in the second detector
to obtain a track projection. The slope of these projections is used
to select forward going protons. A typical slope distribution with a
narrow peak related to forward protons is shown in fig.4a.

Two projections each having at least 5 out of 10 hit layers are
combined to a spatial track. A scatter plot of track points in the
middle plane between both subdetectors is shown in fig.4b. Only very
few fake tracks due to misidentified track projections or ambiguous
combinations in multi-track events can be seen outside the sensitive
detector area.

All spatial tracks inside the sensitive detector area are used to form 
global tracks for each pair of vertical and horizontal FPS stations.
Before this step the track points have to be corrected for the 
detector positions. The large distance between two FPS stations
allows to measure the slopes of global tracks with an accuracy of a
few $\mu$rad.

For a minimum multiplicity of 5 fiber hits the probability to find a
local track projection is 86 \%. This results in a reconstruction
efficiency of about 50 \% for protons passing both vertical stations.
In the horizontal stations this efficiency is smaller due to
the lower layer efficiency (section 4.3).

\subsection{Fiber Detector Efficiency and Resolution}
The performance of the fiber detectors is described by the layer
efficiency, which is defined as the probability that a fiber layer
indicates a hit if it is traversed by a charged particle. The layer
efficiency depends on the light yield, the attenuation of the
scintillating and light guide fibers, and losses in the readout chain
and the hit finding algorithm. Another source of inefficiency is the
dead material between neighbouring fibers. Assuming an effective fiber
core diameter of 900 $\mu$m the geometrical layer efficiency is 86 \%
for a fiber pitch of 1050 $\mu$m.

The hit multiplicity of reconstructed tracks is used to evaluate the
layer efficiencies. For a typical run range the layer efficiencies of
all FPS stations are shown in fig.5. In the vertical stations the values
vary between 50 \% and 80 \% with an average around 65~\%. In the
horizontal stations the layer efficiencies are slightly lower. They
range from 30 \% to 70 \% with an average around 50 \%. The lower
values in the horizontal stations can be explained by the lower
quantum efficiency and gain of the MCPM-124, as given in table 1.
In all FPS stations a degradation of the layer-efficiencies between
5 \% and 10~\% per
year was observed. Possible reasons are the aging of
PSPMs and fiber detectors, intensified by the synchrotron radiation
and beam induced background during injection and beam steering in the
HERA tunnel.
To maintain the quality of the FPS stations fiber detectors with too
low layer-efficiency and PSPMs with a gain degradation were exchanged
during the accelerator shut-down periods.

The spatial resolution of the fiber detectors is determined by the
overlap region of hit fibers in subsequent layers. Due to the
210 $\mu$m staggering of fibers in neighbouring layers a theoretical
resolution of 60 $\mu$m can be obtained if all fibers are 100 \%
efficient. In practice this resolution is detoriated by inefficient
fibers, noise hits and the dead material between the fibers.

For a horizontal station the measured spatial resolution as a function
of the hit multiplicity is shown in fig.6. The average
value of the resolution is 150 $\mu$m, in good agreement with
results from prototype measurements at the DESY electron test beam [8].
Due to the different combinations of overlapping fibers a particular
hit multiplicity results in overlap regions of different sizes.
This effect propagates into the variance of the resolution which 
starts for
all multiplicities around100 $\mu$m and ranges up to 300 $\mu$m for
the lowest multiplicities.

The spatial detector resolution has to be compared with the
uncertainty of the proton trajectory due to vertex smearing and beam
divergence at the interaction point. Depending on the proton energy
these sources give an additional uncertainty between 50 $\mu$m and 150
$\mu$m at the positions of the FPS stations. The Coulomb scattering
of a proton passing a FPS station results in an uncertainty of similar
size in the stations further downstream. Due to these inevitable
contributions, the total resolution cannot be significantly improved
by a better spatial resolution of the fiber detectors.

\subsection{Energy and Angle Reconstruction}
The energy reconstruction is based on the knowledge of the optics
of the proton beam line [10]. Since only dipoles and quadrupoles are
placed between the interaction point and the FPS stations the
deflections in the horizontal and vertical projections are decoupled.
This allows the energy of the scattered proton to be reconstructed
independently in both projections.

The coordinate system in the following description is defined by a
horizontal $X$-axis, a vertical $Y$-axis and a $Z$-axis pointing in beam
direction. The intercept $X$ and slope $X'$ of global tracks in the
horizontal projection are related to the energy E and the scattering
angle $\Theta_x$ of the proton at the interaction point by two linear
equations:

\begin{eqnarray}
        X  = a_{x}(E) + b_{x}(E) \cdot \Theta_x  \\
        X' = c_{x}(E) + d_{x}(E) \cdot \Theta_x 
\end{eqnarray}
A corresponding set of equations holds for the vertical projection.
The transfer functions $a_{x}(E),b_{x}(E),c_{x}(E)$ and $d_{x}(E)$ are
determined from Monte Carlo simulations of forward protons at the
reference positions $Z$ = 85 m and $Z$ = 72 m for the vertical and
horizontal stations respectively. 

While the reconstruction in the horizontal projection has a unique
solution for energy and angle, the vertical projection has two
solutions in most cases. In the vertical stations a large fraction of
these double solutions
has unphysically large scattering angles and can thus be rejected. For
the remaining tracks the solution with the energy closer to the energy
found in the horizontal projection is accepted. The final proton
energy is the weighted average of the energies from both projections.

The energy spectrum of scattered protons measured with the vertical
FPS stations during 1996 data taking with 820 GeV proton beam energy
is shown in fig.7a. It begins around 500 GeV and drops sharply above
750 GeV. The upper edge of the energy spectrum reflects the fact that
in both vertical station the closest distance of approach to the
circulating proton beam is about 4 mm. 

The errors of the reconstructed energies are shown in fig.7b. 
The mean error is 6 GeV at 700 GeV proton energy and decreases to 2
GeV for protons at 500 GeV.

To evaluate the absolute energy scale error, the FPS
measurements were compared with predictions of the proton energy from
the central H1 detector. Events with a high photon virtuality, where
the hadronic final state is well contained in the central detector,
were used for this purpose. Based on a sample of 17 events we estimate
an energy scale error of 10 GeV [14].

The reconstructed spectra of the polar angles $\Theta_x$ and
$\Theta_y$ of scattered protons are shown in fig.7c,d.  The
acceptance in the vertical stations is limited to values 
$|\Theta_{x,y}| < 0.4$~mrad 
corresponding to transverse momenta $p_T <
300$ MeV/c. The mean error of the $\Theta_x$ measurement is 5 $\mu$rad
independent from the proton energy, while the mean error of $\Theta_y$
increases from 5 $\mu$rad at 500 GeV up to 100 $\mu$rad at 700 GeV.

\subsection{Calibration of the Detector Positions}
The transfer functions in (1) and (2) are calculated with respect to
the nominal beam orbit. Since the actual beam orbit varies for
different proton fills, a fill-dependent calibration of the detector
positions is necessary before the track parameters can be used for the
energy reconstruction. This calibration is based on the comparison of
global tracks with a Monte Carlo sample using the nominal beam optics.

Fig.8a shows the scatter plot of intercepts $X$ and slopes $X'$ of
global tracks in the horizontal projection at the reference position
$Z$ = 85 m before the calibration. Superimposed are the lines of
constant energy and scattering angles assuming the nominal interaction
point and beam optics. Certain combinations of $X$ and $X'$ are forbidden,
but this region is partly occupied by uncalibrated tracks. This
discrepancy is based on the difference between nominal and actual beam
orbit and can be improved by additional offsets for slopes and intercepts
of global tracks - the calibration constants $\Delta X$ and $\Delta X'$.
They are determined in a maximum likelyhood fit minimizing the number
of tracks in the forbidden region. Fig.8b shows the scatter plot after
the calibration with much less tracks in the forbidden region.

In the vertical projection the same scheme is applied. Due to the
shape of the scatter plot a unique solution of the calibration
constants $\Delta Y$ and $\Delta Y'$ can only be found if the energy
measurement in the horizontal projection is used as an additional
constraint.

In the horizontal FPS stations the calibration can be done in a similar
way. In addition, the diffractive photo-production of $\rho$-mesons [15],
where the final state is completely measured in the central detector
and the FPS, offers an independent method to check the energy scale.

\section{Summary and Outlook}
For the H1 experiment at the HERA collider a spectrometer was built to
measure forward protons with energies greater than 500 GeV in the
angular range below 1 mrad with respect to the proton beam direction.
Such protons escape the central detector through the beam pipe and can
be detected at a large distance from the interaction point where their
positions deviate a few millimeters from the circulating proton beam.

The FPS consists of two vertical stations at 81 m and 90 m which
approach the beam from above and two horizontal stations at 64 m and
80 m which approach the beam from the outer side of the HERA ring.

The main components of the FPS are fiber detectors located in Roman
Pots, which can be moved close to the proton orbit. The detectors
consist of 5 staggered layers of scintillating fibers of 1 mm
diameter. The scintillating fibers are spliced to light guide fibers
which transmit the signals to position-sensitive photo-multipliers.
For triggering each FPS station is equipped with four planes of
scintillator tiles.

The fiber detectors in the vertical FPS stations have a spatial
resolution of 150 $\mu$m and a typical layer-efficiency of 65 \%. This
allows the proton trajectory to be reconstructed through both stations
for about 50 \% of the triggered events. In the horizontal FPS
stations the spatial resolution is as good as in the vertical
stations, but the layer-efficiency is slightly less, typically 50 \%.

The local track elements are combined into global tracks for the pairs
of vertical and horizontal FPS stations. The parameters of these
global tracks are used to evaluate the energy and scattering angle of
the proton at the interaction point. The deflection of scattered
protons in the horizontal and vertical projection is decoupled and the
energy can be measured twice. For 820 GeV proton beam energy the
vertical stations measure scattered protons in the energy range
between 500 GeV and 750 GeV. The corresponding error varies between 2
GeV and 6 GeV for the lowest and highest energies respectively, with
an additional energy scale error of 10 GeV.

The kinematic range of the horizontal FPS stations is complementary to
that of the vertical stations. The horizontal stations give access to
the diffractive region with 
$\xpom <~0.1$, where $\xpom = 1 - E'/E_p$ is the fractional energy of
the exchange. This is illustrated in fig.9.

Until 1999 an integrated luminosity of about 15 pb$^{-1}$ was
collected with the vertical stations and 5 pb$^{-1}$ with the
horizontal FPS stations. First physics results for the structure
function with a leading proton have been published from the 1995 data
with
the vertical stations [16]. The semi-inclusive structure function
$F_2^{LP(3)}$ with leading protons in the kinematic range 580 GeV $<
E' < $ 740 GeV and $p_T <$ 200 MeV was measured [17]. In another
analysis the photo-production cross section with leading protons and
two jets was evaluated and compared to theoretical predictions [18].

\vspace*{7cm}
\bf
\begin{center}
  Acknowledgement
\end{center}
\rm

The technical help provided by the workshops of the DESY laboratories
at Hamburg and Zeuthen is greatfully acknowledged. In particular we
thank the technicians H.J.Seidel and P.Pohl. The continuous assistance
of the HERA machine, survey and vacuum groups is essential for the
successful operation of the Roman Pot devices.  We thank D.P.Johnson
(Brussels), B.Stella (Rome) and J.Zsembery (Saclay) for their help in
the early phase of this project. This project was supported by the
INTAS-93-43 grant.

\newpage

\begin{center}
\bf
References
\end{center}
\rm

\begin{itemize}
  
\item[[1]] H1 Collaboration, I.Abt et al., Nucl.Instr.and
    Meth.A386 (1997) 310, ibid.348.
\item[[2]] H1 Collaboration, T.Ahmed et al., Phys.Lett.B348 (1995) 681;\\
   H1 Collaboration, S.Aid et al., Nucl.Phys.B463 (1996) 3;\\
   H1 Collaboration, S.Aid et al., Nucl.Phys.B468 (1996) 3;\\
   H1 Collaboration, S.Aid et al., Nucl.Phys.B472 (1996)3, ibid.32.
\item[[3]] T.Regge, Nuovo Cimento 14 (1959) 951;\\
   G.Chew and S.Frautschi, Phys.Rev.Lett.7 (1961) 394;\\
   A.Kaidalov, Phys.Rep.50 (1979) 157;\\
   K.Goulianos, Phys.Rep.101 (1983) 169;\\
   A.Donnachie and P.Landshoff, Phys.Lett.B296 (1992) 227.
\item[[4]] G.Alberi and G.Goggi, Phys.Rep.74 (1981) 1;\\
   G.Chew, S.Frautschi and S.Mandelstam, Phys.Rev.126 (1962) 1202;\\
   A.W.Thomas and C.Boros, ADP-98-79/T346,hep-ph/9812264v2, 1999.
\item[[5]] G.Ingelmann and P.Schlein, Phys.Lett.B152 (1985) 256;\\
   A.Donnachie and P.Landshoff, Nucl.Phys.B303 (1988) 634.
\item[[6]] ZEUS Collaboration, M.Derrick et al., Phys.Lett.B356 (1995) 129;\\
   H1 Collaboration, C.Adloff et al., Z.Phys.C76 (1997) 613;\\
   H1 Collaboration, ''Measurement of the diffractive structure function
   $F_2^{D3}$ at low and high $Q^2$ at HERA'', contr. paper to the
   29th Int.Conf.on High Energy Physics ICHEP'98, Vancouver, Canada, 1998.
\item[[7]] U.Amaldi et al., Phys.Lett.43B (1973) 231;\\
   R.Battiston et al., Nucl.Instr.and Meth.A238 (1985) 35;\\
   A.Brandt et al., Nucl.Instr.and Meth.A327 (1993) 412;\\
   F.Abe et al., Phys.Rev.D50 (1994) 5518;\\
   ZEUS Collaboration, J.Breitweg et al., Eur.Phys.J.C2 (1998) 246.
\item[[8]] J.B\"ahr et al., DESY preprint 92-176, 1992;\\
   J.B\"ahr et al., DESY preprint 93-200,93-201, 1993;\\
   J.B\"ahr et al., Nucl.Instr.and Meth.A330 (1993) 103;\\
   J.B\"ahr et al., Nucl.Instr.and Meth.A371 (1996) 380.
\item[[9]] J.B\"ahr et al., DESY-Zeuthen preprint 95-01, 1995.
\item[[10]] D.C.Carey, ''The Optics of Charged Particle Beams'', Harwood
   Academic Publishers, New York, 1987;\\
   K.G.Steffen, ''High-Energy Beam Optics'', Interscience Monographs and
   Texts in Physics and Astronomy, Vol.17, John Wiley and Sons, New
   York, 1995;\\
   K.Wille, ''Physik der Teilchenbeschleuniger und
   Synchrotronstahlungsquellen'', Teubner,Stuttgart, 1992;\\
   J.Ro{\ss}bach and P.Schm\"user, ''Basic course on accelerator
   optics'', DESY preprint M-93-02, 1993.
\item[[11]] M.Beck et al., Nucl.Instr.and Meth.A381 (1996) 330;\\
   T.Nunnemann, Thesis MPIH-V7-1999, University of Heidelberg, 1998.
\item[[12]] C.Zorn, Nucl.Phys.B32 (1993) 377;\\
   J.B\"ahr et al., DESY preprint 99-079, 1999, physics/9907019.
\item[[13]] The filter algorithm corrects the measured amplitudes
   $S_i$  by subtracting the weighted amplitudes of the direct and
   diagonal neighbours according to:\\
\vspace*{-10mm} 
\begin{center}
\begin{footnotesize}
$\hat{S_i} = 1.025 
\left(S_i - 0.15 \sum\limits_{dir.}S_j
          - 0.10 \sum\limits_{diag.}S_j  \right)$
\end{footnotesize}
\end{center}

\normalsize
\item[[14]] H1 Collababoration, ''The Forward Proton Spectrometer of H1'',
   contr.paper pa17-025 to the 28th Int.Conf. on High Energy Physics
   ICHEP'96, Warsaw, Poland, 1997.
\item[[15]] H1 Collaboration, C.Adloff et al., Z.Phys.C75 (1997) 607;\\
   H1 Collababoration, C.Adloff et al., DESY 99-010, 1999,
   hep-ex/9902019, subm.to Eur.Phys.J.C.
\item[[16]] H1 Collababoration, C.Adloff et al., Eur.Phys.J.C6 (1999) 587.
\item[[17]] B.List, Thesis, University of Hamburg, 1996.
\item[[18]] C.Wittek, Thesis, University of Hamburg, 1997.

\end{itemize}  

\newpage

\pagestyle{empty}

\bf
\noindent
Table 1:
\rm
Typical parameters of the position-sensitive photo-multipliers H4139-20
and MCPM-124 used in the vertical and horizontal FPS stations,
respectively. 

\vspace*{0.5cm}

\begin{tabular}{|l|l|l|}
\hline
Characteristics & H4139-20& MCPM-124 \\
\hline
   photo-cathode material &     bi-alkaline  &       multi-alkaline \\
                window   &     glass        &      fiber optic     \\
                size     &     $40 \times 40$ mm$^2$  &     25 mm diam. \\
\hline
   quantum efficiency  &       typ. 20 \%   &       typ. 15 \% \\
   at 400 nm           &                    &                   \\
\hline
   dynode structure    &       proximity mesh  &   micro-channel plates \\
          stages       &       16              &   2  \\
\hline 
   anode pixels        &       8 x 8         &     10 x 10 + 4 x 6 \\
   pixel pitch         &       5.08 mm       &     2.2 mm \\
   pixel size          &       4 mm diam.    &     1.5 $\times$ 1.5 mm$^2$ \\
\hline
   max.voltage         &       2.7 kV        &     2.9 kV \\
   gain                &       $10^6$ at 2.5 kV  &   3 $\times 10^5$ at 2.8 kV\\
\hline
   pulse time          &       2.7 nsec rise  &    2.5 nsec total \\
\hline
   uniformity          &        1 : 3  &       1 : 3  \\

   average cross talk   &       1 \%      &          1 - 2 \% \\
\hline
\end{tabular}

\vspace*{1cm}

\newpage

\bf
\noindent
Table 2:
\rm
Typical parameters of the trigger photo-multipliers XP1911 and R5600
used in the vertical and horizontal FPS stations, respectively.

\vspace*{0.5cm}

\begin{tabular}{|l|l|l|}
\hline
   Characteristics       &            XP1911 &           R5600 \\
\hline
   photo-cathode material &     bi-alkaline   &     bi-alkaline \\
                window   &     glass         &     glass \\
                size     &     15 mm diam.   &     8 mm diam. \\
\hline
   quantum efficiency    &     20 \%    &      20 \% \\
   at 400 nm             &                   &                  \\  
\hline
   dynode structure      &     linear focused &    metal channel \\
          stages         &     10             &    8  \\
\hline 
   max.voltage           &     1.9 kV         &    1.0 kV  \\
   gain                  &     $10^6$ at 1.2 kV  &   $10^6$ at 0.8 kV \\
\hline
   pulse time            &     2.3 nsec rise  &    0.65 nsec rise \\
\hline
\end{tabular}

\newpage

\noindent
\bf
Table 3 :
\rm
The horizontal and vertical profile of the 820 GeV proton beam at the
interaction point and at the positions of the FPS stations in terms of
standard deviations $\sigma_x$ and $\sigma_y$ of a Gaussian
parametrization. For the emittance the value 17$\pi$  mm $\cdot$ mrad
was assumed.

\vspace*{0.5cm}

\begin{center}
\begin{tabular}{|l|l|l|}

\hline
         z / m  &    $\sigma_{x}$/mm  &   $\sigma_{y}$/mm \\
\hline
           0    &      0.17     &       0.05 \\
          64    &      2.49     &       0.82 \\
          80.5  &      1.86     &       0.25 \\
          90    &      1.38     &       0.22 \\
\hline

\end{tabular}
\end{center}


\newpage

\begin{center}
\bf
Figure Captions
\end{center}
\rm

\begin{description}
  
\item[Fig.1] A schematic view of forward beam line a) in the
  horizontal b) in the vertical projection  indicating the 10 $\sigma$
  beam envelope, the main magnets and the FPS stations at 64 m, 80
  m, 81 m  and 90 m.
\item[Fig.2] A horizontal FPS station a) with fiber detectors in
  parking position and a vertical FPS station b) showing the fiber
  detector end faces and the scintillator planes of the trigger counters.
\item[Fig.3] The electronic scheme : the PSPM and trigger PMT signals
  are transmitted via optical links to the VME master controllers - in
  the other direction control signals (HERA clock, pipeline enable
  , fast clear) are sent to the FPS stations.
\item[Fig.4] The local tracks in a horizontal FPS station:
  a) a slope spectrum with the peak signal of forward proton tracks,
  b) spatial track points with the shape of the detector area.
\item[Fig.5] The layer efficiencies of the fiber detectors a,b) in the
  vertical stations during 1996 data taking and c,d) in the horizontal
  stations during 1998 running.
\item[Fig.6] The spatial resolution of the fiber detectors in the
  horizontal stations in dependence on the multiplicity of fiber hits.
  The bars describe the variance according to the different sizes of
  the fiber overlap regions for combinations with the same hit
  multiplicity.
\item[Fig.7] The energy and angular spectra measured with the vertical
  FPS stations in 1996  a) the energy E obtained from the measurements 
  in both projections, b) the error $\Delta$E in dependence on
  the energy E, and the polar angles c) $\Theta_x$, and d) $\Theta_y$.
\item[Fig.8] The scatter plots of intercepts X and slopes dX/dZ of
  global tracks in the vertical stations a) before and b) after the
  calibration of the detector positions. The full lines correspond 
  constant energies from 420 GeV to 820 GeV, the dashed lines are for
  constant angles from $-0.7$ mrad to $+0.7$ mrad.
\item[Fig.9] The geometrical acceptance in dependence on the
  fractional momentum $\xpom$ of the exchange for protons passing a)
  the vertical and b) the horizontal FPS stations.
\end{description}


\newpage
\psfig{file=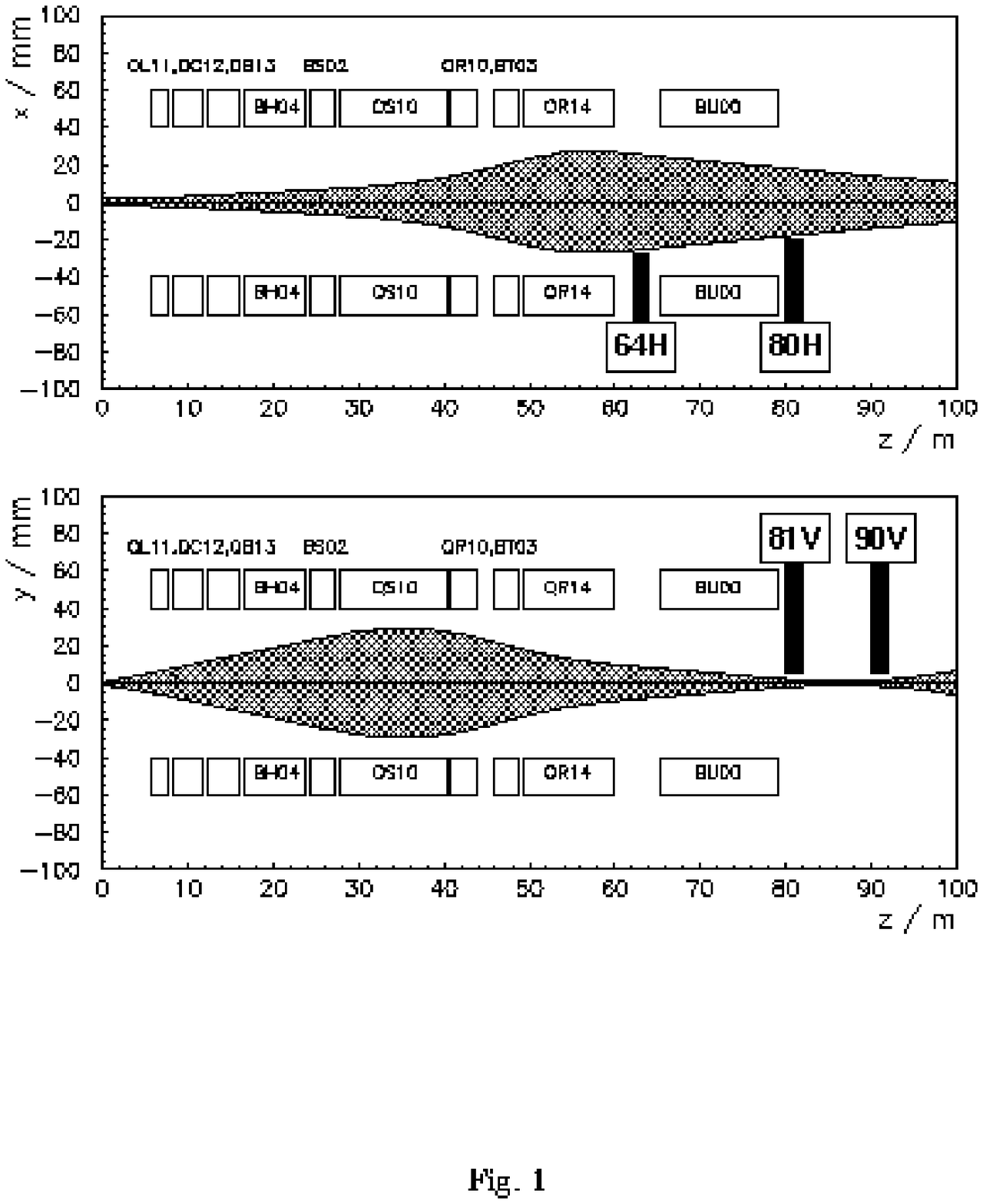}

\newpage
\psfig{file=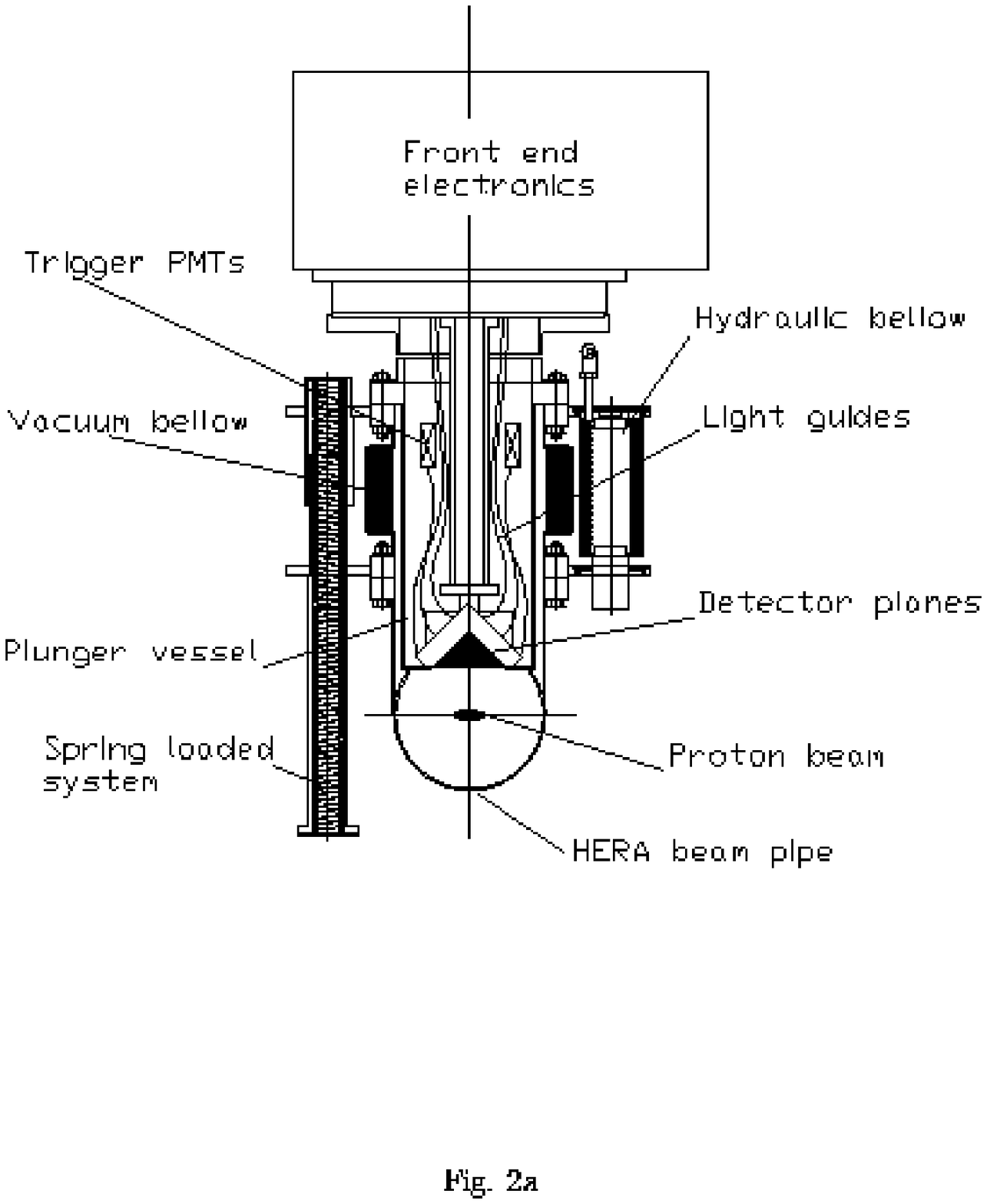}

\newpage
\psfig{file=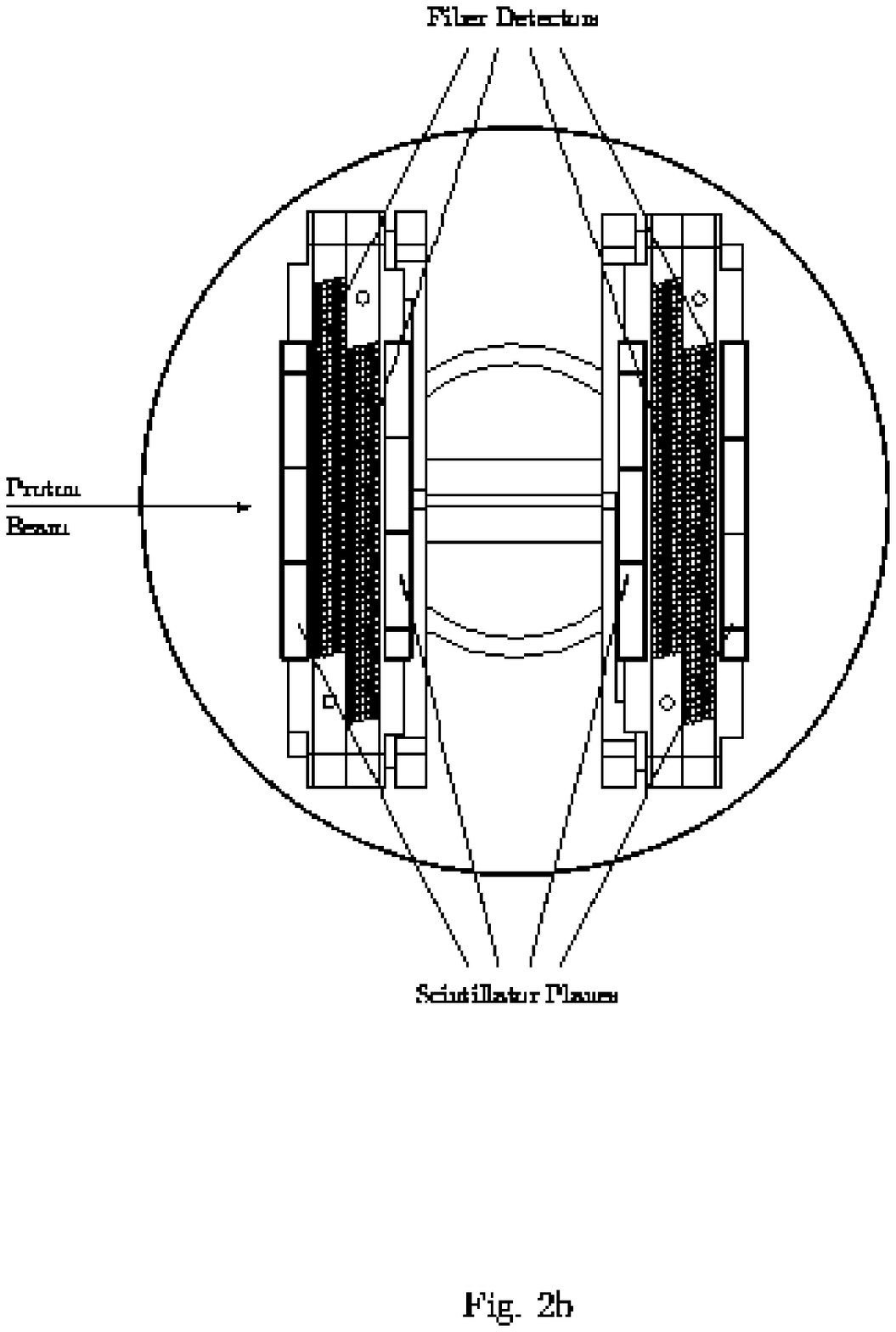}

\newpage
\psfig{file=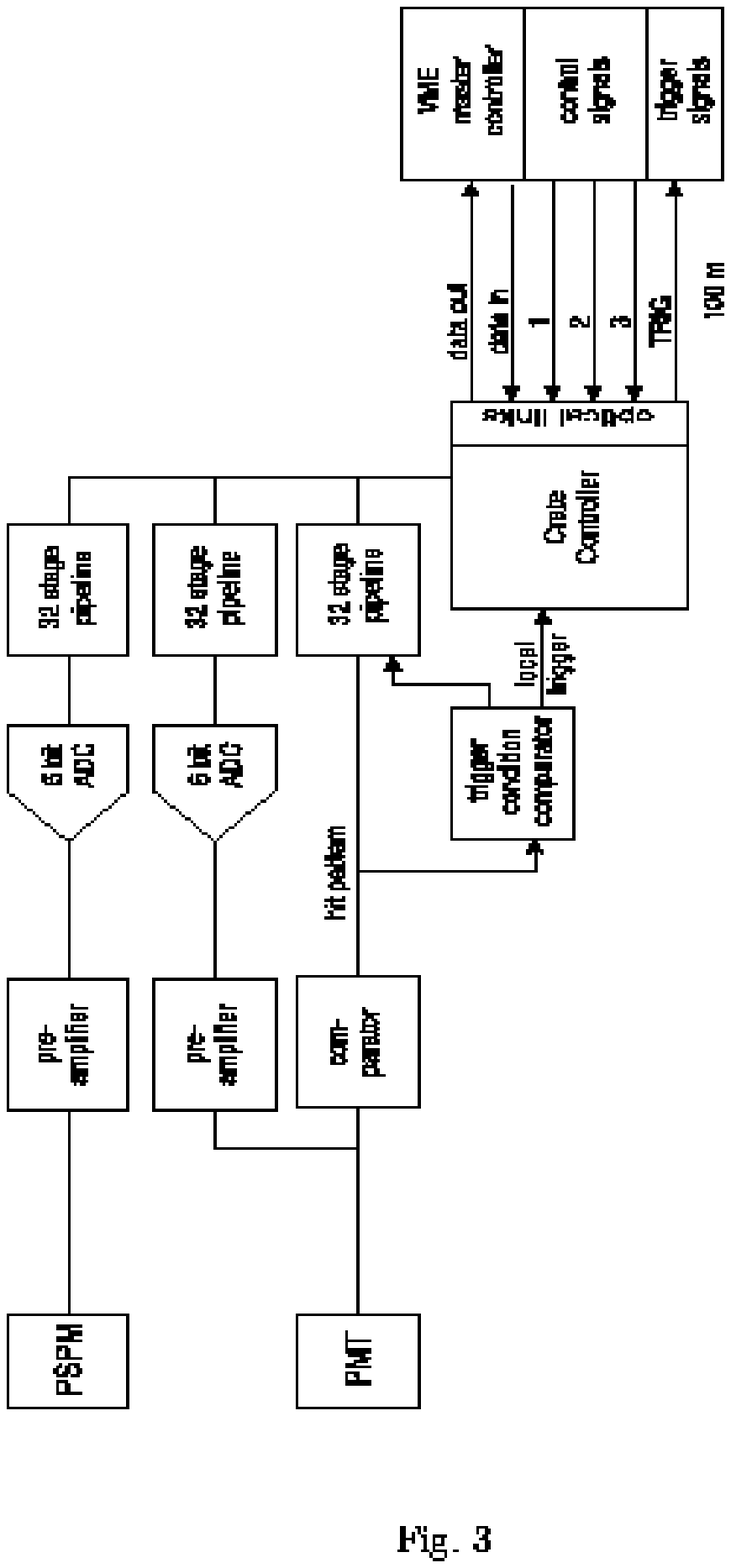}

\newpage
\psfig{file=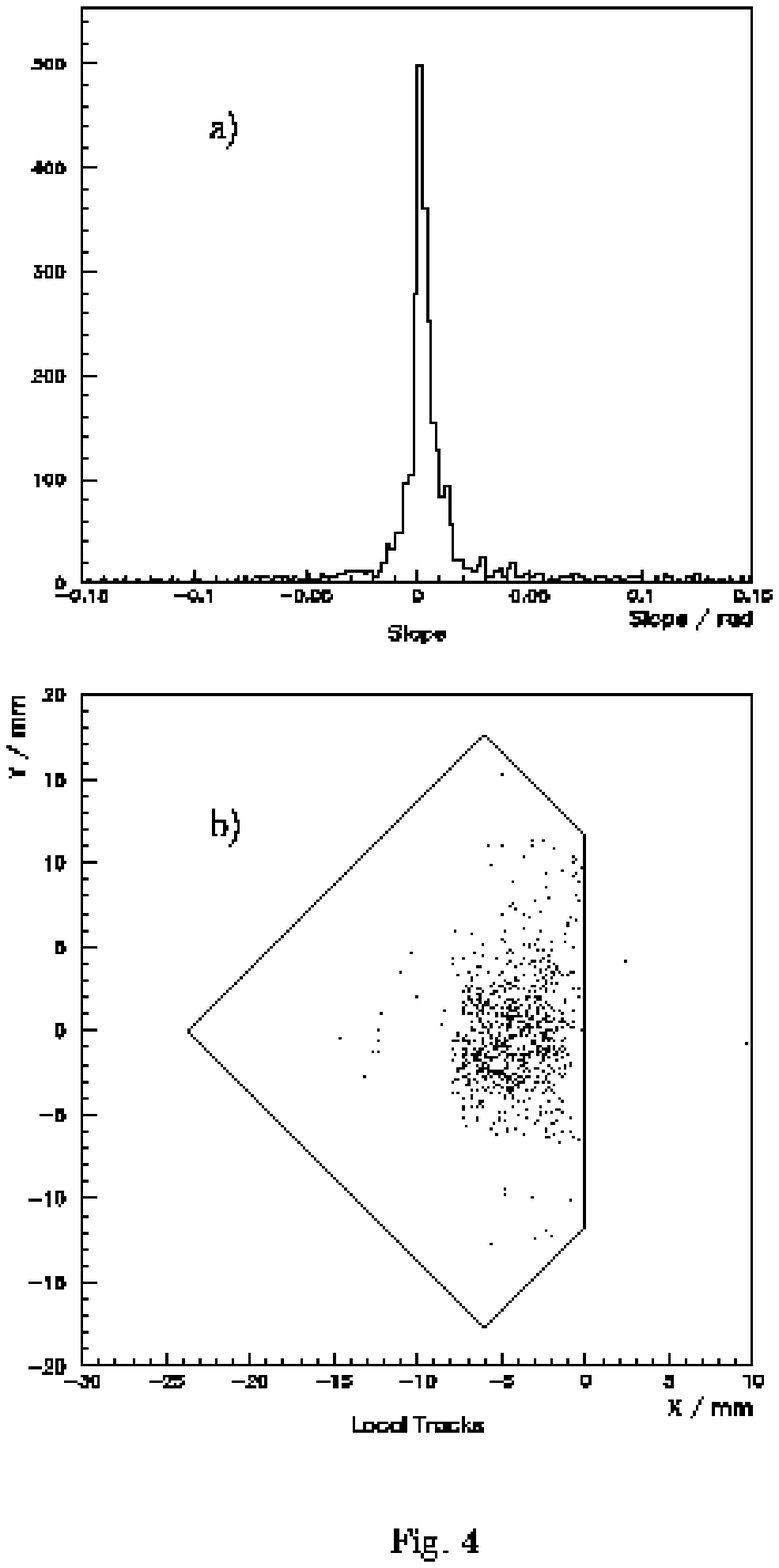}

\newpage
\psfig{file=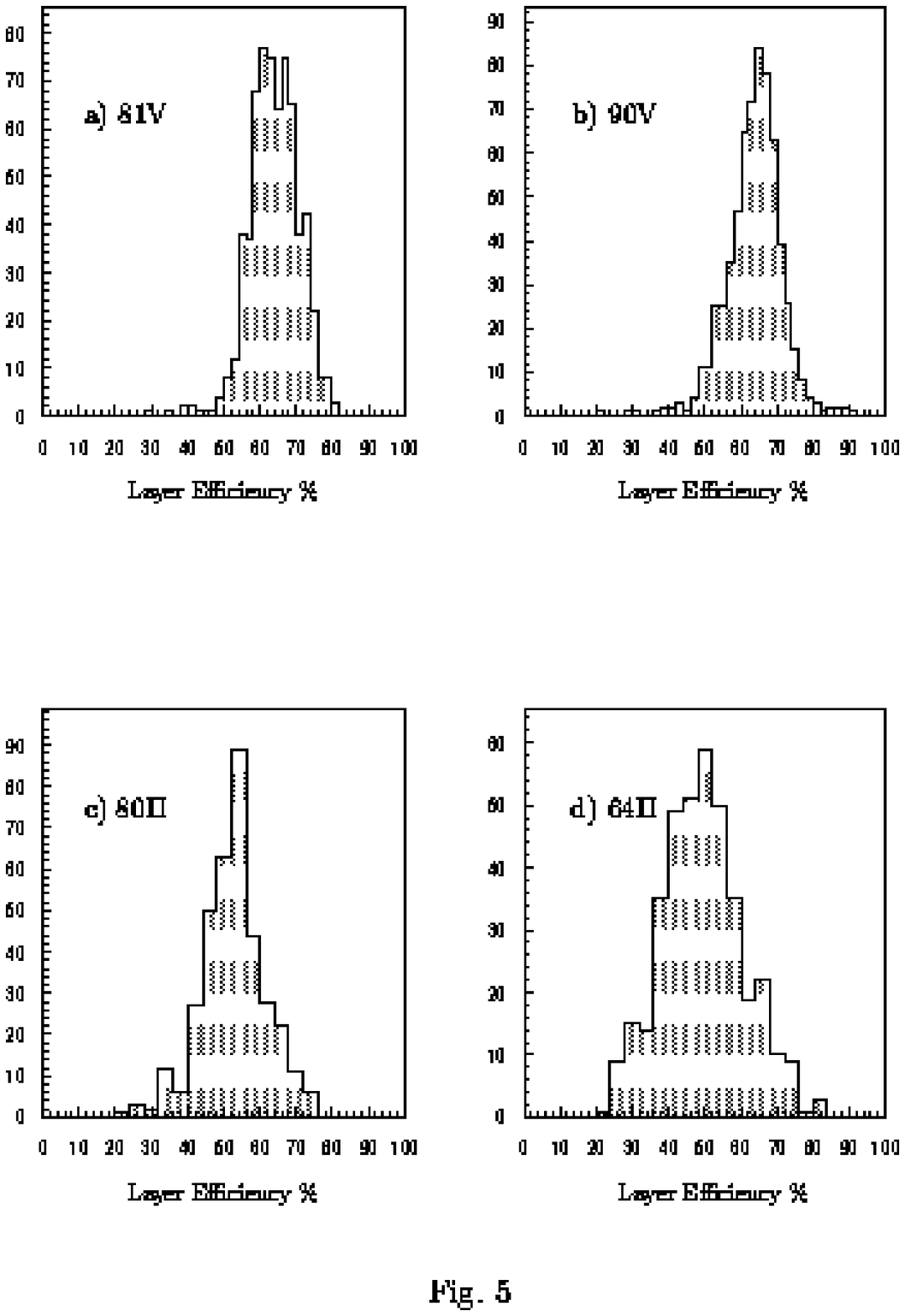}

\newpage
\psfig{file=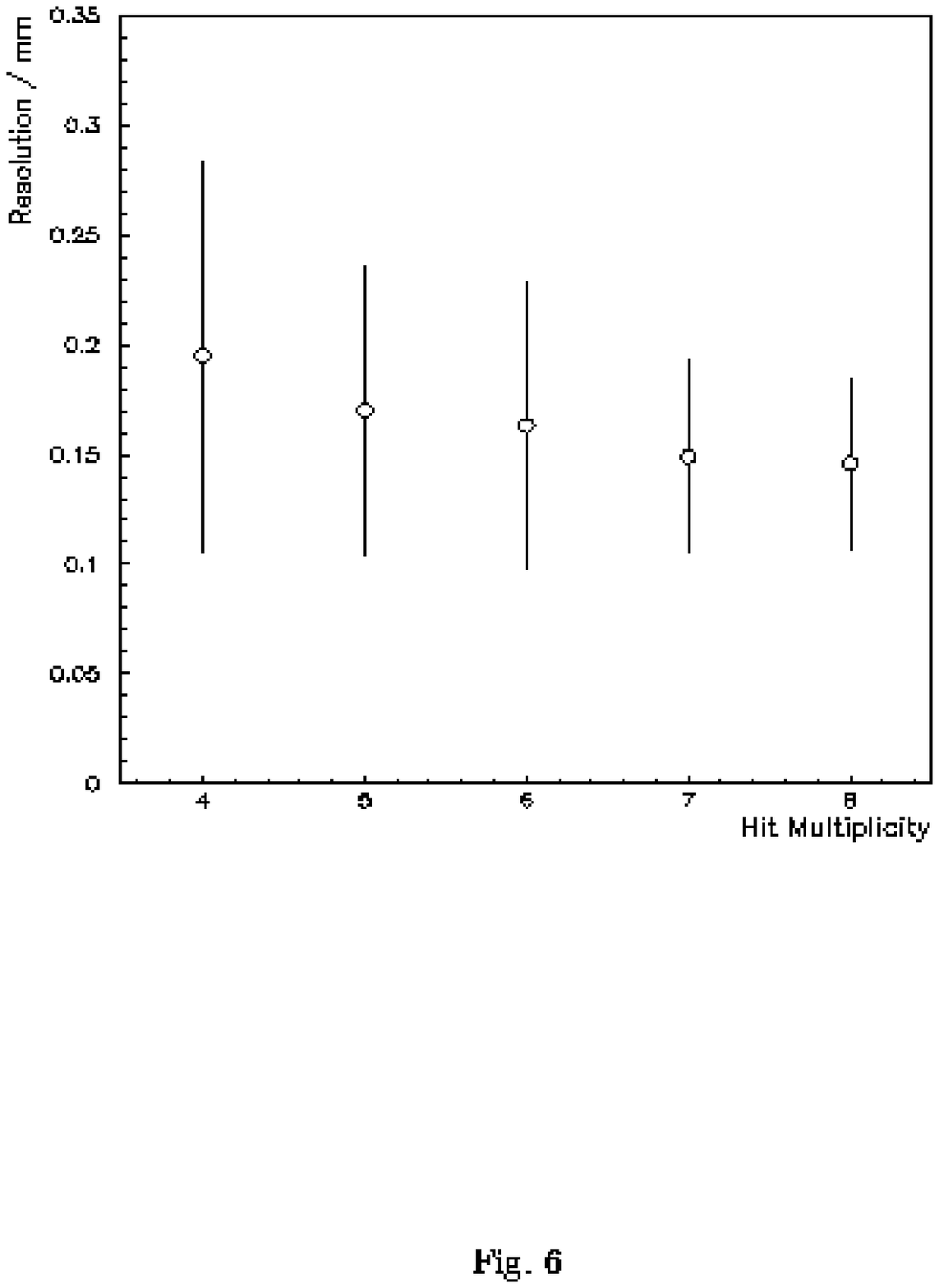}

\newpage
\psfig{file=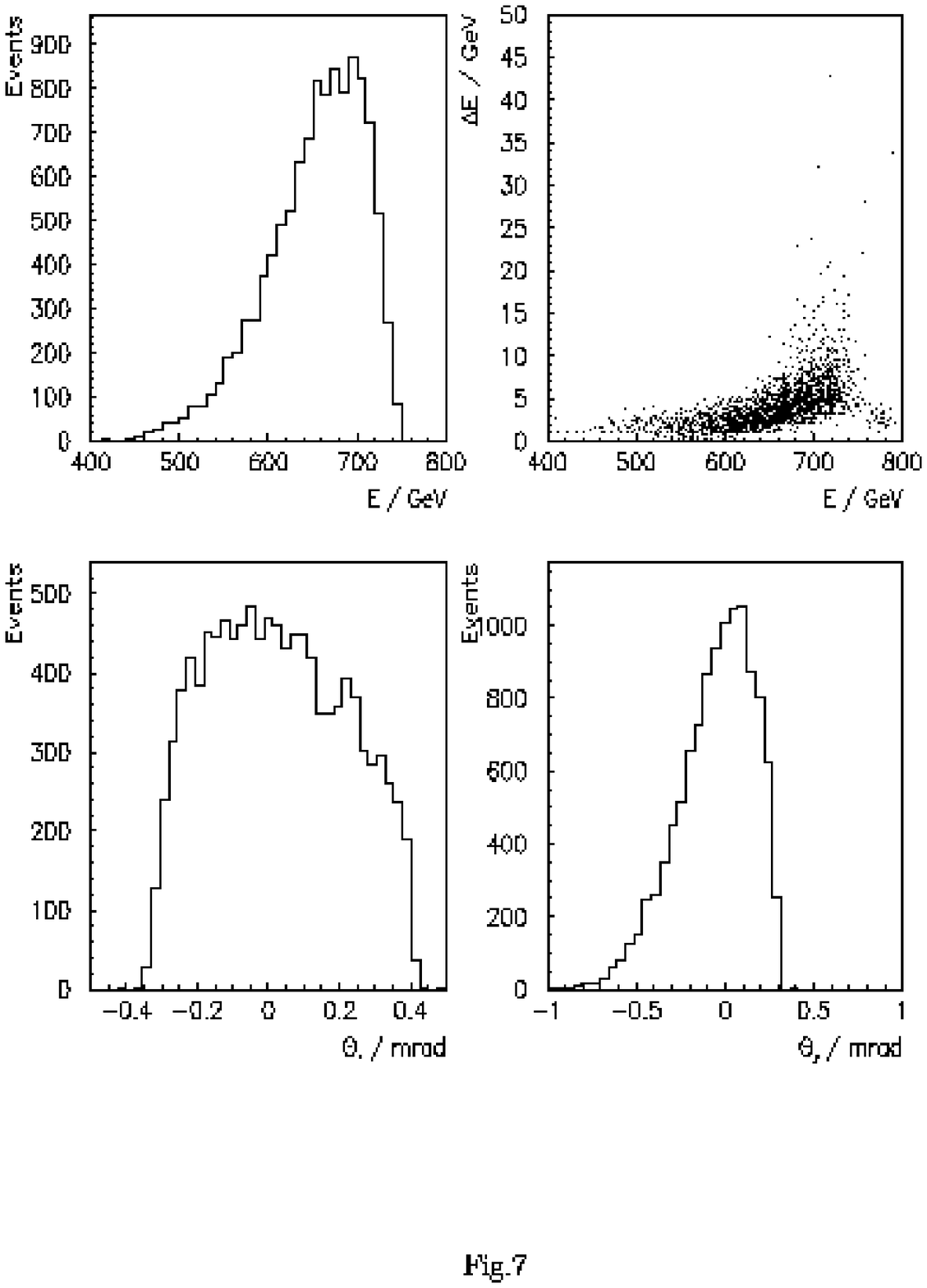}

\newpage
\psfig{file=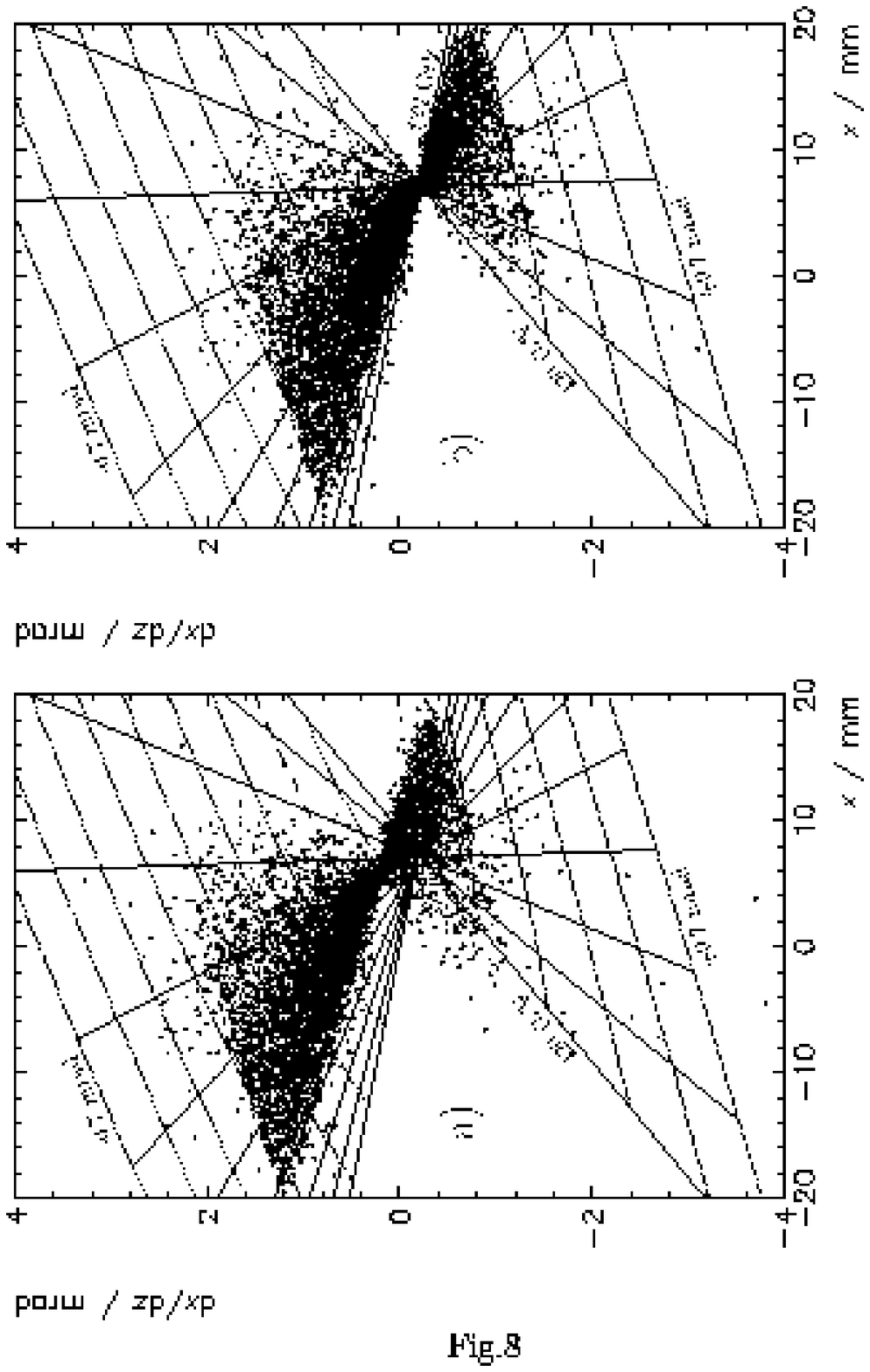}

\newpage
\psfig{file=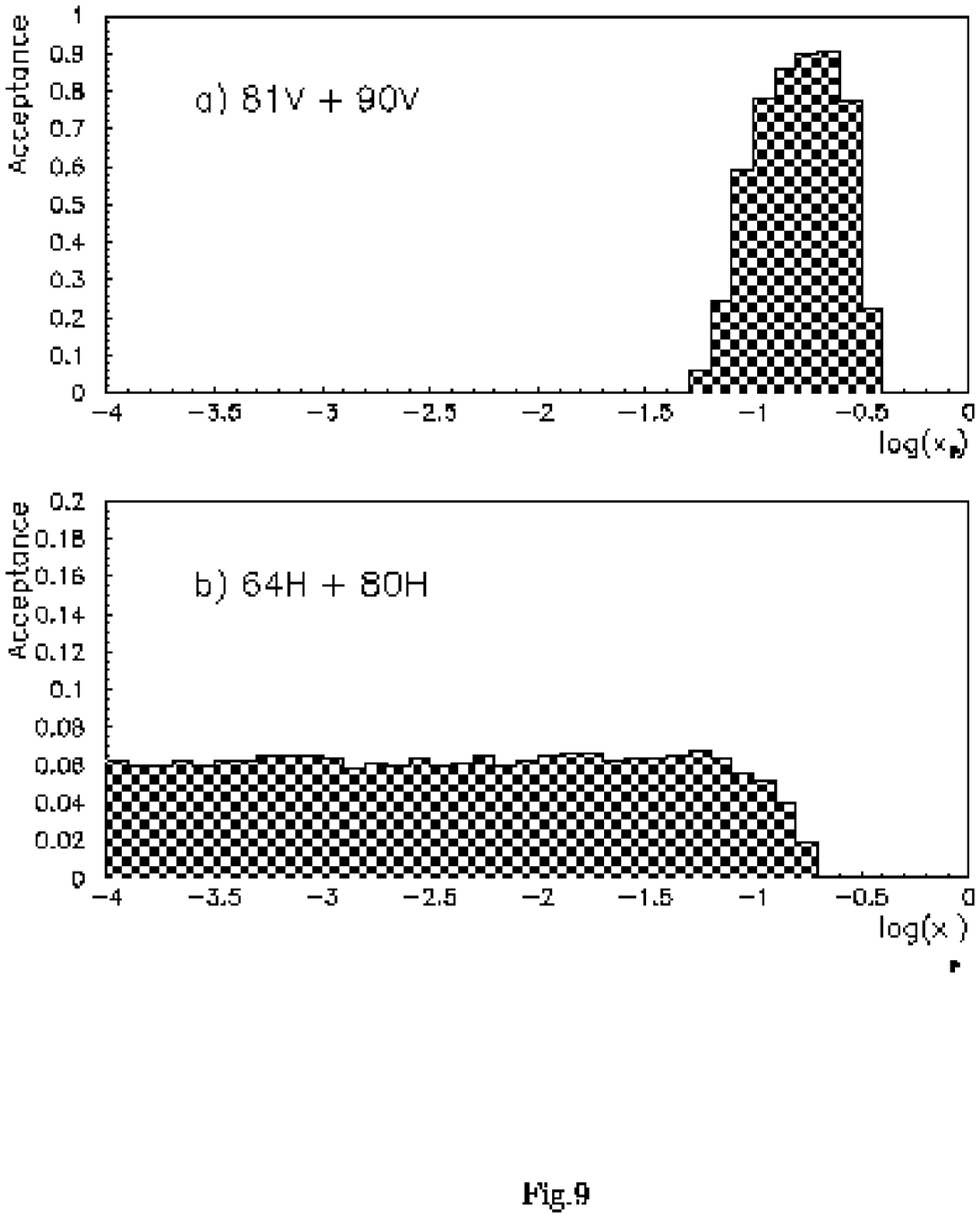}

\end{document}